\documentclass[aps,preprint,letterpaper,amssymb,showkeys,10pt]{revtex4}
\usepackage{epsfig} 
 
\newtheorem{lem}{Lemma}[section]
\begin{document}
\title{Thermodynamics of the Farey Fraction Spin Chain}
\author{Jan Fiala and Peter Kleban}

\affiliation{LASST/SERC and Department of Physics \& Astronomy, University of Maine, Orono, ME 04469}
\email{jan.fiala(at)umit.maine.edu, kleban(at)maine.edu; fax:  (207)-581-2255; phone:  (207)-581-2258}
\date{\today}
\begin{abstract} 
We consider the Farey fraction spin chain, a one-dimensional model defined on (the matrices generating) the Farey fractions. We extend previous work on the thermodynamics of this model by introducing an external field $h$. From rigorous and more heuristic arguments, we determine the phase diagram and phase transition behavior of the extended model.  Our results are fully consistent with scaling theory (for the case when a ``marginal" field is present) despite the unusual nature of the transition for $h=0$.

\end{abstract}

\maketitle

\section{Introduction}\label{Intro}
Phase transitions in one-dimensional systems are unusual, essentially because, as long as the interactions are of finite range and strength, any putative ordered state at finite temperature will be disrupted by thermally induced defects, and a defect in one dimension is very effective at destroying order. Despite this, there are many examples of one-dimensional systems that do exhibit a phase transition. The Farey Fraction Spin Chain (FFSC) \cite{K-O} is one such case, which has attracted interest from both physicists and mathematicians \cite{FK, Ka-O, Pe}.  (Since this work uses some methods that may be unfamiliar to the latter, we include a paragraph at the end of this section outlining our results from a mathematical viewpoint.)

One can define the FFSC as a periodic chain of sites with two possible spin states ($A$ or $B$) at each site. This model is rigorously known to exhibit a single phase transition at temperature $\beta_c = 2$ \cite{K-O}.  The phase transition itself is most unusual.  The low temperature state is completely ordered \cite{K-O,C-Kn} . In the limit of a long chain, for $\beta > \beta_c$, the
system is either all $A$ or all $B$.  Therefore the free energy $f$
is constant and the magnetization $m$ (defined via the difference in the number
of spins in state $A$ vs.~those in state $B$) is completely saturated over
this entire temperature range.  Thus, even though the system
has a phase transition at finite temperature, there are no
thermal effects at all in the ordered state. The same thermodynamics occurs in the Knauf spin chain (KSC) \cite{K, C-K, K-o, G-K}, to which the FFSC is closely related.

\begin{figure}[h]
\centering\includegraphics[width=11.0cm]{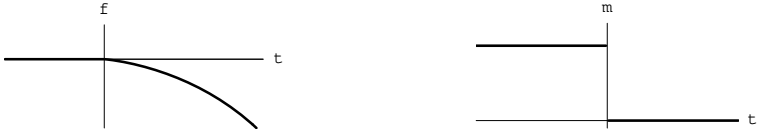}\caption{Free energy and magnetization vs reduced temperature
$t=\frac{\beta_c}{\beta}-1$}
\label{FIG}
\end{figure}

At temperatures above the phase transition (for $\beta < \beta_c$),
 fluctuations occur, and $f$ decreases with $\beta$.
  Here the system is paramagnetic, since (when the external field vanishes, see below)
 there is no symmetry-breaking field.  Thus as the temperature increases $m$
 jumps from its saturated value in the ordered phase to zero
 in the high-temperature phase \cite{C-K,C-Kn} (see Fig.~\ref{FIG}).  (The KSC behaves similarly.)   

One-dimensional models with long-range ferromagnetic interactions \cite{Aiz1,Aiz2} are known to exhibit a discontinuity in $m$ at $\beta_c$, but in these cases the jump in $m$ is less than the saturation value.

The discontinuity in $m$ might suggest a first-order
 phase transition, but in our model the behavior with temperature is different. In
previous work, we proved that as a function of temperature, $f$ exhibits a
 second-order transition, and the same transition occurs in the KSC and the ``Farey tree" multifractal model \cite{FK}.  

In beginning the research reported here, our motivation was to see whether the phase transition in the FFSC, which seems to mix first- and second-order behavior, is consistent with scaling theory.  Indeed, as will be made clear, it is, in the ``borderline" case when a marginal variable is present.   In order to see this, 
 we extend the definition of the FFSC to include a finite external field $h$. We then determine the phase diagram and free energy as a function of $\beta$ and $h$, using both rigorous and renormalization group (RG) analysis.  

In the following, section \ref{DEF} defines the model. Then, in section \ref{Free} we prove the existence of the free energy $f$ with an external field, and evaluate $f$ for temperatures below the phase transition.  In section \ref{RG} we employ renormalization group arguments to find the free energy and phase diagram for temperatures above the phase transition.  Section \ref{KDP} considers a simple model that has very similar thermodynamics but is completely solvable. Section \ref{SUM} summarizes our results. In the Appendix we present some arguments needed to prove the existence of $f(\beta,h)$ in section \ref{Free}.

Since our results may be of interest to mathematicians who are unfamiliar with some of the physics employed herein, we pause to include a description of them from a more mathematical point of view. Section \ref{DEF} defines the model and the quantities of interest.  More specifically, the partition function $Z_N$ is a two-parameter weighted sum over the (matrices defining the) Farey fractions, and the free energy $f$ then follows from the limiting procedure defined in (\ref{freeE}).  The main goal of our work is to find the analytic behavior of $f$ as a function of the real parameters $\beta$, the inverse temperature (so $\beta > 0$ is implicit), and $h$, the external field.  Regions of parameter space for which $f$ is analytic are (thermodynamic) phases, and the lines of singularities that separate them are phase boundaries.  In section \ref{Free} we prove that $f(\beta,h)$ exists, and compute it exactly at low temperature (for $\beta>\beta_c$), which constitutes part of the ordered phase. Section \ref{RG} uses renormalization group methods to determine $f$ at high temperatures (for $\beta$ near $\beta_c$ and $\beta<\beta_c$).  Since this method is not rigorous, from a mathematical point of view the results should be regarded as conjectures.  The main conclusions are the form of the free energy in the high-temperature phase (\ref{ffd}, \ref{sf}), the equation for the phase boundary (\ref{b}, \ref{bc}) and the change in magnetization $m=-\partial f/\partial h$ (\ref{cm}) and entropy $s=\beta^2 \, \partial f/\partial \beta$ across the phase boundary. We also find that the ordered phase, with $f= \mp h$, extends to $\beta<\beta_c$ when $h$ is sufficiently large (see Fig.~\ref{P}).  Section \ref{FSS} gives predictions for the behavior of $Z_N$ as $N \to \infty$ near the second-order point ($\beta = \beta_c$ and $h=0$). This is related to some work in number theory, but unfortunately not yet directly. Section \ref{KDP} examines an exactly solvable model with certain similarities to the FFSC.

\section{Definition of the model}\label{DEF}
The FFSC consists of a periodic chain of $N$ sites with two possible spin states ($A$ or $B$) at each site.  The interactions are long-range, which allows a
 phase transition to exist in this one-dimensional system. 
Let the matrices
\begin{equation}\label{M}
M_N:=\prod_{i=1}^{N}A^{1-\sigma_i}B^{\sigma_i},\qquad \sigma_i\in \{0,1\},
\end{equation}
where $A:=\left({1\atop 1}{0\atop 1}\right)$ and $B:=\left({1\atop 0}{1\atop 1}\right)$ and the dependence of $M_N$ on $\{ \sigma_i\}$ has been suppressed.
The energy of a particular configuration with $N$ spins in an external field $h$ is given as
\begin{equation}\label{en}
E_N:=\ln (T_N)+h\left ( 2\sum_{i=1}^{N}\sigma_i - N  \right )\quad {\rm with} \quad T_N:={\rm Tr}(M_N).
\end{equation}
Thus our partition function is
\begin{equation}\label{pf}
Z_N(\beta, h)=\sum_{\{\sigma_i\}}{\rm Tr}(M_N)^{-\beta}e^{ -\beta h\left ( 2\sum_{i=1}^{N}\sigma_i - N  \right )}.
\end{equation}
This definition extends the Farey fraction spin chain model to non-vanishing external field $h$.
Given the nature of the low-temperature $h=0$ system, it is natural to introduce $h$ in this way. 

The free energy is defined as
\begin{equation}\label{freeE}
f(\beta,h) :=\frac{-1}{\beta} \lim_{N\rightarrow\infty}\frac{\ln
Z_N(\beta,h)}{N}.
\end{equation}
The existence of the free energy $f(\beta,h)$ follows from simple bounds using $f(\beta,0)$ (see section \ref{Free} below).

The definition of the FFSC is somewhat unusual. The partition function is given in terms of the energy of each possible
 configuration, rather than via a Hamiltonian. In fact, there is no known way
 to express the energy exactly in terms of the spin variables \cite{K-O}.  Further,
 numerical results indicate that when one does, the Hamiltonian has all
 possible even interactions (and they are all ferromagnetic), so an 
explicit Hamiltonian representation, even if one could find it, would
be exceedingly complicated. 

Note that for $h=0$ there are two ground states with energy $E=\ln 2$. The other $2^N-2$ states have energy
$\ln N \le E \le N c$, where $c$ is a constant.
Therefore the difference between the lowest excited state energy and the ground state 
energy diverges as $N\to\infty$. 

The phase transition in this system \cite{K-O} occurs in the following way.  Divide the partition function into two terms, one due to the two ground states, and the other (call it $Z'$), due to the remaining $2^N-2$ states. The system remains in the ground states, and $Z' \to 0$ as $N \to \infty$, until the 
temperature is high enough that $Z'$ diverges with  $N$. In section \ref{KDP} 
we examine a simple model that also exhibits this feature, but is completely solvable. 

Our results also apply to the KSC, which has
the same thermodynamics as the FFSC model at $h=0$ (see \cite{FK}). An external field may be included in the KSC in exactly the same way as described above for the FFSC.  The ``Farey tree" model of Feigenbaum et.~al.~\cite{F} also has the same free energy, but it is not clear how to incorporate a field $h$.  Our finite-size results (see section \ref{FSS}) do apply when $h=0$, however.

\section{Free energy with an external field}\label{Free}
In this section we show rigorously that $f(\beta,h)$ exists and that
\begin{equation}\label{fh}
f(\beta, h)=-|h|,
\end{equation}
for $\beta > \beta_c$.

For $h>0$ it is easy to see (from (\ref{pf})) that 
\begin{equation}\label{ine}
2^{-\beta} e^{\beta hN}<Z_N(\beta, h) <Z_N(\beta, 0) e^{\beta hN}.
\end{equation}
Using the definition of the free energy then gives
\begin{equation}\label{feb}
-h \ge f(\beta, h)\ge f(\beta,0)-h,
\end{equation} 
where $f(\beta,h)$ is understood to be defined via (\ref{freeE}). Now $f(\beta,0)$ is rigorously known to exist \cite{K-O}.
In addition, we know that $f(\beta, 0)=0$ for $\beta \geq \beta_c$ \cite{K-O}, which implies (\ref{fh}) for $h>0$ ($h<0$ follows similarly).

To see that $f(\beta,h)$ exists for the range $0\le \beta <\beta_c$ we proceed as follows (actually, our argument applies for all $\beta \ge 0$). 
We first show that $\left | \frac{\log Z_{N+1}}{N+1}-\frac{\log Z_{N}}{N}\right |\to 0$ as $N\to \infty$.  The result then follows by use of (\ref{ine}). Now 

$$\left | \frac{N\log Z_{N+1}-N\log Z_{N}-\log Z_{N}}{N(N+1)}
\right |\le \left | \frac{\log Z_{N+1}/Z_{N}}{N+1}\right |+\frac{1}{N+1}\left | \frac{\log Z_{N}}{N}\right |,$$
and we see by (\ref{ine}) and the existence of $f(\beta,0)$ that
the second term $\frac{1}{N+1}\left | \frac{\log Z_{N}}{N}\right |\le \frac{K}{N+1}$ for some finite constant $K$.
In the appendix we show  that $2^{-\beta}e^{-\beta |h|}\le\frac{Z_{N+1}}{Z_N}\le 2 e^{\beta |h|}$ which completes our proof of the existence of the free energy for
all $\beta \ge 0$ and $h\in {\mathbb R}$.

We also know rigorously that $f(t,0)\sim c\frac{t}{\ln t}+..$, where $c>0$, $t=\frac{\beta_c}{\beta}-1$, for $t>0$ (see Fig.~\ref{FIG}).
It follows that $f(t, h)$ must have at least one 
singularity between the regions with low and high temperatures, i.e.~a phase transition from the ordered to the high-temperature phase. 

Since we can not calculate $f(\beta, h)$ exactly for $\beta < \beta_c$ (except for $h=0$ and $\beta \to \beta_c$ ), we use another method, in the next section, to 
examine the thermodynamics.

\section{ Renormalization group analysis} \label{RG}
\subsection{Mean field theory}
In mean field theory one assumes that there is an expansion of the free energy of the form 
\begin{equation}\label{MF}
f_{MF}=a+btM^2+uM^4- g hM+\ldots \ ,
\end{equation} 
where $M$ is the magnetization and the ``constants" $a$, $b$, $u$ and $g$ are weakly dependent on the reduced temperature $t$ (defined at the end of section \ref{Free}) and external field $h$. Note that $u > 0$ is required for stability, and $b>0, g \ge 0$ in the high-temperature phase. (The possibility that $g=0$ is ruled out below.)

Minimizing (\ref{MF}) with respect to $M$, one obtains the free energy and magnetization in mean field approximation. Explicitly
\begin{enumerate}
\item for $t>0$ and $h\ne 0$ the magnetization $$M_0\sim \frac{1}{6}\left [ \frac {u}{gh} +\left (\frac{2bt}{3gh}\right )^3 \right ]^{-\frac {1}{3}}$$
(note the limiting cases $M_0 \sim 0$ for $h=0$ and $M_0\sim 1/6(gh/u)^{1/3}$ for $t=0$)
\item for $t<0$ and $h\ne 0$, but $h$ sufficiently small, the magnetization 
 $$M_0\sim \left ( \frac{b|t|}{2u} \right )^{\frac {1}{2}}+\frac{gh}{4b|t|}$$
(however when 
$\left (\frac{gh}{2u}\right )^2+4\left( \frac{bt}{6u} \right )^3>0$, $M_0$ is given by the $t>0$ formula).
We include this second case only for completeness. Since our system is completely saturated at low temperatures this result is not employed in our analysis. 
\end{enumerate}
In the following we use the first result in an RG analysis.
\subsection{Renormalization group analysis}

We assume two relevant fields ($t$ and $h$) and one marginal field ($u$).  These assumptions are reasonable, since our model has an Ising-like ordered state, the interactions are (apparently) all ferromagnetic, and there is a logarithmic term in the free energy. 

The infinitesimal renormalization group transformation for the singular part of the free energy is
\begin{equation}\label{RF}
f_s(t,h,u)=e^{-d\ell}f_s(t(\ell),h(\ell),u(\ell)).
\end{equation} 
Because of the marginal field $u$, the analysis is somewhat more complicated than otherwise.  We follow the treatment of Cardy (\cite{Ca}, see also Wegner \cite{We}). The RG equations take the  form
\begin{eqnarray}
du/d\ell&=&-xu^2+\ldots \label{RE1} \\ 
dt/dl&=&y_tt-z_tut+\ldots \label{RE2} \\
dh/dl&=&y_hh-z_huh+\ldots ,\label{RE3}
\end{eqnarray} 
where we keep only the most important terms. The omitted terms are either higher order or go to zero more 
rapidly with $\ell$ than those included.
From (\ref{RE1}) we find (note $t=t(0)$, $h=h(0)$, $u=u(0)$)
\begin{equation}\label{u}
u(\ell)=\frac{u(0)}{1+xu(0)\ell}.
\end{equation} 
Both $t$ and $h$ have the same functional form, namely
\begin{equation}\label{t}
\ln(t(\ell_0)/t(0))=y_t\ell_0-\frac{z_t}{x}\ln [1+xu(0)\ell_0]
\end{equation} 
and
\begin{equation}\label{h}
\ln(h(\ell_0)/h(0))=y_h\ell_0-\frac{z_h}{x}\ln [1+xu(0)\ell_0], 
\end{equation}
where $\ell_0$ is such that $t(\ell_0)=O(1)$ or  $h(\ell_0)=O(1)$.
From (\ref{t}) we can write 
\begin{equation}\label{l}
\ell_0 \sim \frac{1}{y_t} \ln \frac{t_0}{t} +\frac{z_t}{xy_t}\ln \left [1+\frac{x}{y_t}\;u\ln\frac{t_0}{t}\right ],
\end{equation} 
where we assume $t_0/t \gg 1$.
This result together with (\ref{RF}) gives us
\begin{equation}\label{f1}
f_s(t,h,u)\sim\left |\frac{t}{t_0} \right |^{\frac{d}{y_t}}\left [1+\frac{x}{y_t}\;u\ln\frac{t_0}{t}\right ]^{-\frac{z_t \;d}{y_t\; x}}
f_s(t(\ell_0),h(\ell_0),u(\ell_0)).
\end{equation} 
Since the free energy on the rhs is evaluated at $\ell _0$, which is far from the critical point,
 it can be calculated from mean field theory. Above the critical temperature ($t>0$)
 with small external field $h$ ($h(\ell_0)\ll t(\ell_0)$) we obtain for the free energy 
\begin{equation}\label{FE}
f_s(t(\ell_0),h(\ell_0),u(\ell_0))\sim a-\frac{3(gh(\ell_0))^2}{16bt(\ell_0)}.
\end{equation} 
The relation between $h(\ell_0)$ and $t(\ell_0)$ follows from (\ref{t}) and (\ref{h}).
Eliminating $h(\ell_0)$ allows us to rewrite (\ref{FE}) as 
\begin{equation}\label{FEt}
f_s\sim a-\left | \frac{t_0}{t} \right |^{2\frac{y_h}{y_t}}h^2
\left [1+\frac{x}{y_t}\;u\ln\frac{t_0}{t}\right ]^{2y_h\left [\frac{z_t}{y_t\; x}-\frac{z_h}{y_h\; x}\right ] }
\left (-\frac{3g^2}{16bt(\ell_0)}\right ).
\end{equation} 
Substituting the result into (\ref{RF}) gives two terms,
\begin{equation}\label{1}
\left |\frac{t}{t_0} \right |^{\frac{d}{y_t}}\left [1+\frac{x}{y_t}\;u\ln\frac{t_0}{t}\right ]^{-\frac{z_t \;d}{y_t\; x}}a,
\end{equation} 
and
\begin{equation}\label{2a}
\left | \frac{t}{t_0} \right |^{\frac{d}{y_t}-2\frac{y_h}{y_t}}h^2
\left [1+\frac{x}{y_t}\;u\ln\frac{t_0}{t}\right ]^{-\frac{z_t \;d}{y_t\; x}+2y_h\left [\frac{z_t}{y_t\; x}-\frac{z_h}{y_h\; x}\right ] }
\left (-\frac{3g^2}{16bt(\ell_0)}\right ). 
\end{equation} 

The first term can be compared with the exact result at $h=0$ (see section \ref{Intro}). It follows that
\begin{equation}\label{com1}
\frac{d}{y_t}=1=\frac{z_t}{x}.
\end{equation} 
The second term gives us the dependence on external field. Eliminating $t(\ell_0)$ instead of $h(\ell_0)$ we obtain
\begin{equation}\label{2b}
\frac{1}{t}\left | \frac{h}{h_0} \right |^{\frac{d}{y_h}+\frac{y_t}{y_h}}
\left [1+\frac{x}{y_t}\;u\ln\frac{t_0}{t}\right ]^{-\frac{z_h \;d}{y_h\; x}-y_t\left [\frac{z_h}{y_h\; x}-\frac{z_t}{y_t\; x}\right ] }
\left (-\frac{3(gh(\ell_0))^2}{16b}\right ). 
\end{equation} 

Equating the two expressions (\ref{2a}) and (\ref{2b}) for the same term in the free energy gives us the RG eigenvalues 
\begin{equation}\label{ytd}
\frac{d}{y_t}=\frac{d}{y_h}=1,
\end{equation} 
where $d$ is the dimensionality of the system. This is of course one for our model, but since none of our results require setting $d=1$ we leave it unspecified. 

Finally we can write down the singular part of the free energy for the high-temperature phase
\begin{equation}\label{fd}
f_s(t,h,u)\sim\left |\frac{t}{t_0} \right |\left [\frac{x}{y_t}\;u\ln\frac{t_0}{t}\right ]^{-1}a
-\frac{h^2}{t}
\left [\frac{x}{y_t}\;u\ln\frac{t_0}{t}\right ]^{1-\frac{z_h}{x}}
\left (\frac{3g^2}{16b}\right ).
\end{equation} 
Since $f<0$ for $h=0$ in this phase, (\ref{fd}) implies that $a<0$. 

For the ordered phase we know rigorously that the free energy has no temperature dependence for $h=0$. 
The spins are all up or all down. When we add an external field it will break the symmetry and all the 
spins will be oriented in the field direction. Thus the free energy at $\ell_0$ is 
\begin{equation}\label{mfo}
f_s(t(\ell_0),h(\ell_0),u(\ell_0))=-|h|(\ell_0).
\end{equation} 
Proceeding as in the derivation of (\ref{f1}) from (\ref{RF}) and (\ref{t}) we get 
\begin{equation}\label{f2}
f_s(t,h,u)\sim\left |\frac{h}{h_0} \right |^{\frac{d}{y_h}}\left [1+\frac{x}{y_h}\;u\ln\frac{h_0}{h}\right ]^{-\frac{z_h \;d}{y_h\; x}}
f_s(t(\ell_0),h(\ell_0),u(\ell_0)),
\end{equation} 
using (\ref{h}),
 (\ref{mfo}) and (\ref{ytd}) then give
\begin{equation}\label{fo}
f_s(t,h,u)\sim-|h|\left [1+\frac{x}{y_h}\;u\ln\frac{h_0}{h}\right ]^{-\frac{z_h}{x}}.
\end{equation} 
Because the magnetization in the ordered state is completely saturated the logarithmic correction must vanish.
Therefore $z_h=0$.

Thus the asymptotic form for the free energy of the high-temperature state is 
\begin{equation}\label{ffd}
f_s(t,h,u)\sim\left |\frac{t}{t_0} \right |\left [\frac{x}{y_t}\;u\ln\frac{t_0}{t}\right ]^{-1}a
-\frac{h^2}{t}
\left [\frac{x}{y_t}\;u\ln\frac{t_0}{t}\right ]
\left (\frac{3g^2}{16b}\right ).
\end{equation} 
We can recast this result more suggestively as 
\begin{equation}\label{sf}
f_s(t,h,u)\sim f_s(t,0,u)-\frac12 h^2\chi(t,0,u) ,
\end{equation}
where $\chi = -\, \partial^2 f/\partial h^2$ is the susceptibility. Note that $\chi \sim 1/f_s$ which is consistent with scaling theory, since (using (\ref{ytd})), $f \sim t^{2-\alpha} = t^{d/y_t} = t$ while $\chi \sim t^{-\gamma} = t^{(d-2 y_h)/y_t} = t^{-1}$. This relation holds regardless of whether we set the dimensionality $d=1$ or not.  In addition, the coefficient of $\frac{t}{\ln t}$ for the free energy at $h=0$ and $t \to 0,\, t>0$ is known exactly \cite{Diss,P-S}, so that the combination of constants $\frac{y_t a}{|t_0| x u}$ may be determined.

The phase boundary is given by the continuity of the free energy.  Now we expect the ordered phase to exist for $\beta < \beta_c$ if $h$ is large enough (this is reflected in the assumption of two relevant fields-if another phase intervened there would be more). Thus one must equate the two expressions for $f$. One finds that the phase boundary between the ordered and high-temperature phase, close to the critical point, follows
\begin{equation}\label{b}
|h|\sim k\frac{t}{\ln t/t_0},
\end{equation} 
where $k=\left \{ \frac{8 b y_t}{3  x u g^2}\left [  1-\sqrt{1+\frac{3 a g^2}{4 b t_0}}\  \right ]\right \}$.
Since $f$ is quadratic in $h$ in the high-temperature phase, there are in general two solutions with $h>0$.  However, the one at larger $h$ is not physical since it gives rise to a magnetization $m>1$ and violates the convexity of the free energy as well, so we employ the other.
\begin{figure}[h]
\centering\includegraphics[width=5.0cm]{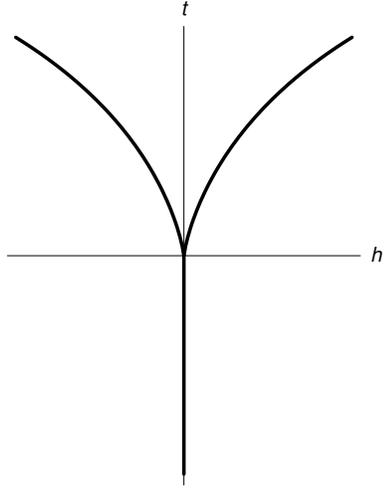}\caption{Phase diagram}
 \label{P}
\end{figure}

In order to find the change in magnetization across the phase boundary we use (\ref{b}) with constants included
\begin{equation}\label{bc}
|h|\sim \frac{-t}{\ln \frac{t_0}{t}}\left \{ \frac{8 b y_t}{3  x u g^2}\left [  1-\sqrt{1+\frac{3 a g^2}{4 b t_0}}\  \right ]\right \}.
\end{equation} 
In arriving at (\ref{bc}), we (as mentioned) chose the root that makes $m<1$ in the high-temperature phase. Note that in the limiting case that $\frac{3 a g^2}{4 b t_0}=-1$, $m=1$ but the two roots coincide.

Now from (\ref{ffd}) 
\begin{equation}\label{md}
m\sim\frac{h}{t}
\left [\frac{x}{y_t}\;u\ln\frac{t_0}{t}\right ]
\left (\frac{3g^2}{8b}\right ).
\end{equation}
Eliminating the external field using (\ref{bc}), and since the magnetization in the ordered phase takes the values $m\sim \pm 1$, we find
\begin{equation}\label{cm}
\Delta m\sim \sqrt{1+\frac{3 a g^2}{4 b t_0}}.
\end{equation}
Note that $t_0$ is a constant of order one and recall that $a<0$, thus on the phase boundary the discontinuity in magnetization is constant (and less than one),
at least close to the second-order point (we argue below that $g=0$ is not possible in this model).
Now we can look at the change in entropy (per site) $s=\beta^2 \, \partial f/\partial \beta$ across the phase boundary.
We get 
\begin{equation}\label{sb}
\Delta s\sim -2 \left [\frac{x}{y_t}\;u\ln\frac{t_0}{t}\right ]^{-1}
\left (\frac{a}{t_0}+\frac{4b}{3g^2} \left [  1-\sqrt{1+\frac{3 a g^2}{4 b t_0}} \right ] \right ).
\end{equation}
These results show that the phase transition is first-order everywhere except at $h=0$.  

In the limiting case when $\frac{3 a g^2}{4 b t_0}=-1$, already mentioned, one finds that both $\Delta m = 0$ and $\Delta s = 0$.  However, it is easy to see that both the susceptibility $\chi$ and the specific heat will have a discontinuity across the phase boundary.

Note that the magnetization change given by (\ref{cm}) exhibits a kind of ``discontinuity of the discontinuity", in that its limiting value as one approaches the second-order point is not the same as its value at that point.  This is not the case for the entropy change, or for these quantities in the model examined in section \ref{KDP}.

Finally, we argue that $g=0$ is not possible in the high-temperature phase.  Since the second derivative of $f$ with respect to $h$ at $h=0$ is proportional to both $g$ and the susceptibility $\chi$, it suffices to demonstrate that $\chi>0$.  It is straightforward to show that $\chi$ is proportional to $\Sigma_{j=1}^{N} \langle s_1 s_j \rangle $ where the spin variables $s_i := 2\sigma_i -1,\quad s_i\in \{-1,1\}$ (cf.~(\ref{M})), and the angular brackets denote a thermal average.  Now the $j=1$ term in this sum is $1$, and due to the ferromagnetic interactions in the spin chain, the remaining terms cannot be negative. Note that this argument is not completely rigorous, since for the FFSC we only have numerical evidence that the interactions are all ferromagnetic.  The KSC, on the other hand, is known to have all interactions ferromagnetic \cite{C-Kn}, so that $\langle s_1 s_j \rangle >0$ follows from the GKS inequalities.

\subsection{Finite-size  scaling} \label{FSS}
We can use our results to make some predictions about finite-size  (i.e.~$N$ large but $N<\infty$) effects on the thermodynamics.
We make the standard assumption that the size of our spin chain is a relevant field with eigenvalue $1$. Of course, since our system has long-range interactions the validity of finite-size scaling may be questioned \cite{Ca},
 but it is still interesting to see the results.
The treatment is the same as in the case of 
the relevant fields $t$ and $h$. The renormalization equation for the inverse size $I := N^{-1}$ is then
\begin{equation}\label{RE4}
dI/dl=I-z_{I}u I+\ldots\ .
\end{equation} 
Thus we get 
\begin{equation}\label{feN}
f_s(t, h,u, N^{-1})\sim \left |  \frac{N_0}{N}\right |^d\left [1+x\;u\ln\frac{N}{N_0}\right ]^{-\frac{z_{I}\,d}{x}}
f_s(t(\ell_0), h(\ell_0),u(\ell_0),N^{-1}(\ell_0)).
\end{equation}
Note that we do not know the ratio $z_I/ x$, however (\ref{feN}) gives the form we should observe. More succinctly, for large $N$, this result predicts that for small $t$ and $h$
\begin{equation}\label{pfu}
\ln Z_N(t,h)\sim N^{1-d}[\ln N]^{-p}.
\end{equation}

There is related work in number theory by Kanemitsu \cite{Kan} (cf.~also \cite{Jap}). This paper studies moments of neighboring Farey fraction differences, which are similar to the ``Farey tree" partition function \cite{F}. At $h=0$, the latter has the same thermodynamics as the FFSC \cite{FK}. However, \cite{Kan} uses a definition of the Farey fractions that, at each level, gives a subset of the Farey fractions employed here, and none of the moments considered corresponds to $\beta =2$ (the point of phase transition). It is interesting that, despite these differences, terms logarithmic in $N$ appear. More specifically, the sum of $m$th (integral) moments of the differences goes as
\begin{equation}\label{mom}
\frac { {(\ln N)}^{\delta_{2,m} }} {N^m} +O \left (\frac {{(\ln N)}^{h(m)}} {N^{m+g(m)}} \right ),
\end{equation}
 for $m \ge 2$, with $g(2)=1$, $g(3)=2$ and $g(m)=3$ for $m \ge 4$, and $h(m)=1$ for $2 \le m \le 4$, $h(m)=0$ for $m \ge 5$.
Now if all the Farey fractions were included (\ref{mom}) would apply to the Farey tree partition function with $\beta=2 \, m$ (cf.~\cite{FK,F}) so that $m \ge 2$ would correspond to $\beta \ge 4$. It would be interesting to determine whether (\ref{mom}) applies to the Farey tree partition function despite this difference, or to extend (\ref{mom}) to $m=1$ to see if it is consistent with (\ref{pfu}). 

\section{1-D KDP model with nonzero external field}\label{KDP}

In this section we consider the one-dimensional KDP (Potassium dihydrogen phosphate) model introduced by Nagle \cite{N}. 
This model's thermodynamics and energy level structure are similar to the Farey fraction spin chain, but it is easily solvable. Comparison of the two models thus sheds some light on the FFSC.  

The KDP model exhibits first-order phase transitions only. The origin of the phase transition is infinite rather than long-range interactions. 
\begin{figure}[h] 
\centering\includegraphics[width=7.0cm]{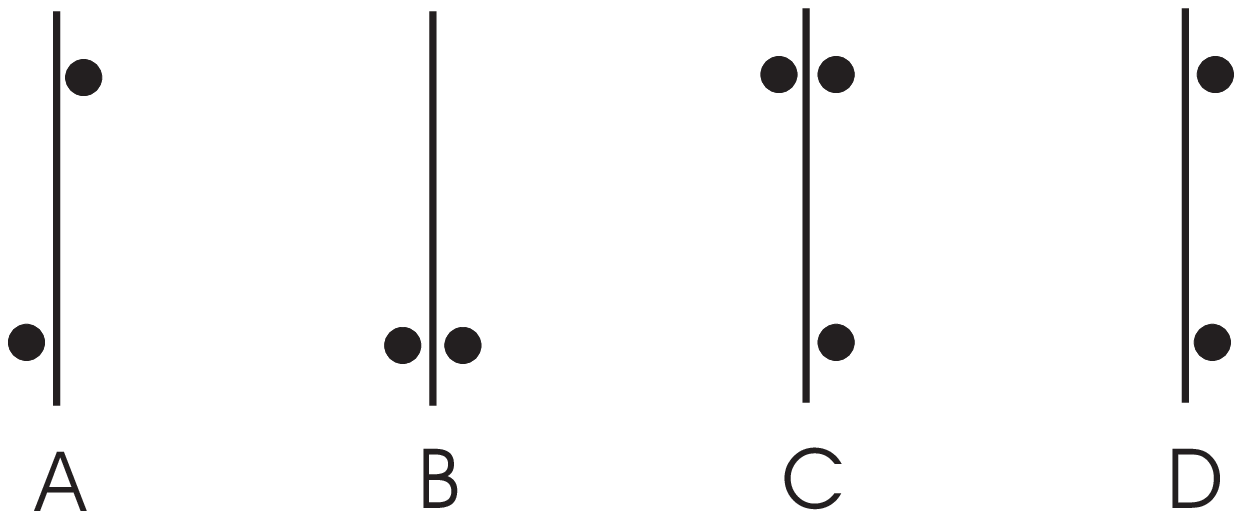}\caption{KDP}
\label{k}
\end{figure}

The one-dimensional geometry of the model is illustrated in Fig.~\ref{k}. 
It consists of $N$ cells, and each cell contains two dots. Each dot represents a proton in a hydrogen bond in the KDP molecule.
 Dots can be on the left or the right side of a cell.
The energy of a neighboring pair of cells depends on the arrangement of dots at their common boundary. 
Only configurations with exactly two dots at each boundary (e.g.~A, B and D in Fig.~\ref{k}) are allowed, any other configuration 
(e.g.~C in Fig.~\ref{k}) has (positively) infinite energy and is therefore omitted. Of the allowed configurations, only two energies occur, $0$ (when there are two dots on the same side of a boundary,
as in Fig.~\ref{k} D) or $\epsilon$ (when the dots are on opposite sides, as in Fig.~\ref{k} A or B).

Let there be $N$ cells in a chain with periodic boundary conditions.
 Then there are two kinds of configurations with finite energy.  In the first type of configuration, each cell has two dots on the same side.
There are two such configurations and the total energy of each is $0$. 
In the second type of configuration, each cell has one dot on the left and one on the right. There are $2^N$ such configurations and the total energy of each is $N\epsilon$. Thus, the partition function is simply
\begin{equation}\label{cp}
Z_N(\beta)=2+2^N\exp{(-\beta N \epsilon)}.
\end{equation}
It follows immediately that $f=0$ for $\beta \epsilon >\ln 2$ and $f=\epsilon-\frac{\ln 2}{\beta}$ for $\beta \epsilon <\ln 2$.
Thus the temperature of the (first-order) phase transition is
 $T_c=\epsilon /(\ln2)$ and there is a latent heat with entropy change $\Delta s=\ln 2$. Clearly, the phase transition mechanism is a simple entropy-energy balance.  At low temperatures, the ground state energy gives the minimal free energy, while in the high-temperature phase the extra entropy of the additional states gives a lower free energy.

Next, define the magnetization $m$ as the number of sides of cells with both dots on one side divided by the number of cells $N$.   
Then $m=1$ for $\beta > \beta_c$ and $m=0$ for $\beta < \beta_c$ (so that
$\Delta m=1$ at the phase transition), just as in the FFSC model.   

\begin{figure}[h] 
\centering\includegraphics[width=5.0cm]{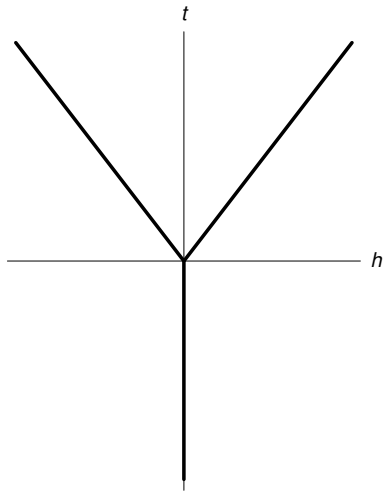}\caption{Phase diagram}
\label{PK}
\end{figure}

Following the above definition of the magnetization, we introduce an external field $h$ by adding an energy $\pm h/2$ to each dot, according to whether it is on the right or left side of the cell. This gives the extra energy of an external field acting along the chain. Then the new partition function has the form 
\begin{equation}\label{KDPZ}
Z_N(\beta, h)=2\cosh (\beta N h) + 2^N \exp (-\beta N\epsilon).
\end{equation}
In the ordered phase $Z\to \exp (\pm \beta N h)$. Thus, the free energy $f=\mp h$, where the plus sign is for $h>0$ and the minus sign for $h<0$, exactly as in the FFSC.
For the high-temperature phase
$Z\to \exp \Bigl [ N(\ln 2-\beta \epsilon) \Bigr ]$ and we get the same free energy as when $h=0$, $f=\epsilon-\frac{\ln 2}{\beta}$. The phase boundary is given by $h= \pm \epsilon t$ (see Fig.~\ref{PK}), where $t = \frac{\beta_c}{\beta}-1$ as before.  Note the resemblance to the FFSC phase diagram (Fig.~\ref{P}).  Here, as $\beta \epsilon \to \ln 2$, $h\to 0$ as it should, while for $\beta \to 0$ the field $h \to \ln 2/ \beta $.
The entropy per site vanishes everywhere in the ordered phase, while for the high-temperature phase $s=\ln2$. 
Thus, this model has a non-zero latent heat and the phase transition is first-order everywhere. 
Note that the change in magnetization is $\Delta m=1$ everywhere along the phase boundary between the ordered state and the high-temperature state.
 
Now for $h=0$, the FFSC has two ground states with all spins up or all spins down
and energy independent of length $N$, just as in the KDP model. Then, in addition, the FFSC has $2^{N}-2$ states with energies between $\ln N  $ and $N c$,
for some constant $c$. On the other hand, the KDP model has just
one energy ($N\epsilon$) for the $2^N$ states corresponding
to the $2^{N}-2$ states of the Farey model. This might suggest that the states with energies close to $\ln N$ are responsible for the logarithmic factor in the Farey free energy, and thus shift the phase transition from first to second-order (for $h=0$). For $h \ne 0$ the energy of the $\ln N$ states is shifted by the field $h$ to order $N$, and the phase transition becomes first-order. However the mechanism of the FFSC phase transition may be more subtle.  The ``density of states" (number of configurations with a given energy) for the FFSC not well-behaved. In fact it is known rigorously that this quantity, summed over all chain lengths, has a limit distribution \cite{Pe}.

Note that the free energy just derived is independent of $h$ in the high-temperature phase. Since this is not what we found for the FFSC, we consider another way to introduce an external field $h$ into the KDP model. 
\begin{figure}[h] 
\centering\includegraphics[width=8.0cm]{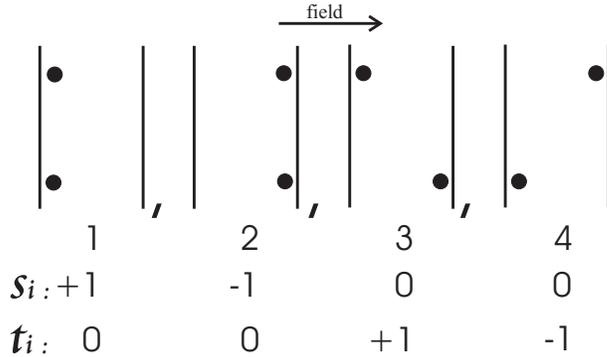}\caption{notation}
\label{KDP2}
\end{figure}

As before we have four different states for each cell. We index them with spin-one variables $t_i$ and $s_i$ ($s_i,\ t_i \in \{ 0,+1,-1\}$ )
in each cell as in Fig.~\ref{KDP2}.
Then the energy (for $h=0$) can be written
\begin{equation}\label{en2}
H_0=\epsilon \sum_{i=1}^{N-1}t_i^2t_{i+1}^2
\end{equation}
(assuming, in the sum, that the infinite energy contributions are omitted). The conditions $s_i + t_i=\pm 1$ and $s_it_i=0$ define the allowed states. We define the magnetization per site as 
\begin{equation}\label{mps}
m=\frac{1}{N}\sum_{i=1}^{N} (s_i+t_i).
\end{equation}
Note that this definition gives a positive (negative) contribution if the upper dot in a given cell is on the right (left). 
(Note also that $m^2=\frac{1}{N^2}\sum_{i=1}^{N} (s_i+t_i)^2+\frac{1}{N^2}\sum_{i\ne j} (s_i+t_i)(s_j+t_j)=\frac{1}{N}\sum_{j=2}^{N} (s_1+t_1)(s_j+t_j)+1/N$.)
Hence we can include an external field as follows
\begin{equation}\label{efi}
H=H_0-h\sum_{i=1}(s_i+t_i)=H_0-hNm.
\end{equation}
Thus
\begin{equation}\label{pf2}
Z(\beta, h)=e^{\beta N h}+e^{-\beta N h}+e^{-\beta \epsilon N}[ 2\cosh (\beta h)]^N,
\end{equation}
and the free energy in high-temperature phase becomes
\begin{equation}\
f(\beta,h)=\epsilon-\frac{\ln(2\cosh(\beta h))}\beta
\end{equation}
or for small $h$
\begin{equation}\
f\sim -t\epsilon -\frac{\ln2}{2 \epsilon (t+1)} h^2,
\end{equation}
with $t=\frac{\beta_c}{\beta}-1$ as above.
The phase boundary is given by 
\begin{equation}\label{phl2}
\beta h=\ln \Big (2 \cosh (\beta h) \Big )-\beta \epsilon.
\end{equation}
For $\beta h \ll 1$ and $h>0$, using $\beta_c=\frac{\ln 2}\epsilon$, this gives 
\begin{equation}\label{sphl2}
h= \epsilon \, t + \frac{\epsilon \ln 2}{2} \,t^2+ O(t^3),
\end{equation}
The phase diagram near the critical point is very close to the previous one (see Fig.~\ref{PK}).
The magnetization in the ordered phase is again independent of temperature, i.e.~$m=\pm 1$. In the high-temperature phase we have $m=\tanh (\beta h)$.
Thus the magnetization change across the phase boundary close to the critical point is $\Delta m= 1-t\ln 2$.
The transition is again first-order, with the entropy change $\Delta s= \ln 2 (1-\frac{ \ln 2}{2}\, t^2)$. Results for $h<0$ follow immediately by symmetry.

\section{Summary and comments}\label{SUM}
 
In this paper, we have extended the definition of the Farey fraction spin chain to include an external field $h$.  From rigorous and more heuristic arguments, we have determined the phase diagram and phase transition behavior of the extended model.  Our results are fully consistent with scaling theory (for the case when a ``marginal" field is present) despite the unusual nature of the transition for $h=0$.  In particular, we find for the renormalization group eigenvalues $y_h=y_t=d$, and for the sub-leading eigenvalues $z_t=x$ and $z_h=0$.  We also examine a completely solvable model with very similar thermodynamics, but for which all phase transitions are first-order.

\section{Acknowlegements}\label{ACK}
We are grateful to J. L. Cardy and M. E. Fisher for useful suggestions.  We also thank an anonymous referee for alerting us to an incompleteness in one of our proofs.  This work was supported in part by the National Science Foundation Grant No. DMR-0203589.
\appendix*
\section{Bounds for $\frac{Z_{N+1}}{Z_N}$}
First we introduce some notation (following \cite{FK}).

We use $r_N^{(n)}:=\frac{n_N^{(n)}}{d_N^{(n)}}$ for
the fractions (called Farey fractions), where
$n$ is the order of the Farey fraction in level $N$. Level $N=0$
 consists
 of the two fractions $\left\{ \frac{0}{1},\frac{1}{1} \right \}$. 
 Succeeding levels are generated by keeping
all the fractions from level $N$ in level $N + 1$, and including new fractions.
  The new fractions at level $N+1$ are defined via
 $ d_{N+1}^{(2n)}:=d_N^{(n)}+d_N^{(n+1)}$ and
 $ n_{N+1}^{(2n)}:=n_N^{(n)}+n_N^{(n+1)}$,
 so that\\
$N=0 \quad  \left \{ \frac{0}{1},\frac{1}{1} \right \}$\\
$N=1 \quad \left \{ \frac{0}{1},\frac{1}{2},\frac{1}{1} \right \}$\\
$N=2 \quad \left \{ \frac{0}{1},\frac{1}{3},\frac{1}{2},\frac{2}{3},
\frac{1}{1} \right \}$, etc.\\[.1cm]
Note that $n=1,\ldots,2^N+1$. When the Farey fractions are defined using matrices
(spin states) A and B, the level $N+1$ is the number of matrices in the chains starting with matrix $A$ and hence the length of the spin chain \cite{K-O}.

Using this notation we can write the partition function (\ref{pf}) restricted to chains starting with $A$
\begin{equation}\label{FF}
Z_N^{A}(\beta,h)=\sum_{n=1}^{2^N}\frac{e^{ -\beta h\left ( 2\sum_{i=1}^{N}\sigma_i - N  \right )}}{(d_N^{(n)}+n_N^{(n+1)})
^{\beta}},\quad \beta
\in \mathbb R.
\end{equation}

Note that the partition function (\ref{pf}) is the sum of $Z_N^{A}(\beta,h)$ and $Z_N^{B}(\beta,h)$, where the $Z_N^{B}(\beta,h)$
is the partition function for chains starting with the matrix $B$.   First we find  bounds for $Z_N^{A}(\beta,h)$ and then prove a lemma
which lets us apply the bounds for $Z_N^{A}(\beta,h)$ to $Z_N^{B}(\beta,h)$ also.

Now, when we go from level $N$ to level $N+1$ we double the number of the terms in the partition function.
Note that for chains starting with the matrix $A$ one half of the terms come from matrix products of the form $AM_{N-1}A$ and the others 
from products $AM_{N-1}B$. It is easy to check that the corresponding traces for given $n\in \{1,\ldots,2^N\}$ are $d_{N+1}^{(2n-1)}+n_{N+1}^{(2n)}$
and $d_{N+1}^{(2n)}+n_{N+1}^{(2n+1)}$, respectively. These traces are multiplied by an $h$ dependent factor
$e^{ -\beta h\left ( 2\sum_{i=1}^{N+1}\sigma_i - N-1  \right )}$ which is simply $e^{\beta h}$ raised to the power $(\#A-\#B)$, 
the number of matices $A$ minus the number of matices $B$ in the particular chain. For the terms from products of the form $AM_{N-1}A$, it follows on using the definition of the Farey fractions that 
$$\frac{e^{ -\beta h\left ( 2\sum_{i=1}^{N+1}\sigma_i - N-1  \right )}}{(d_{N+1}^{(2n-1)}+n_{N+1}^{(2n)})
^{\beta}}=\frac{e^{ -\beta h\left ( 2\sum_{i=1}^{N}\sigma_i - N  \right )+\beta h}}{(d_N^{(n)}+n_N^{(n)}+n_N^{(n+1)})
^{\beta}}\le\frac{e^{ -\beta h\left ( 2\sum_{i=1}^{N}\sigma_i - N  \right )}}{(d_N^{(n)}+n_N^{(n+1)})
^{\beta}}e^{\beta |h|}$$
and, similarly,  for $AM_{N-1}B$
 $$\frac{e^{ -\beta h\left ( 2\sum_{i=1}^{N+1}\sigma_i - N-1  \right )}}{(d_{N+1}^{(2n)}+n_{N+1}^{(2n+1)})
^{\beta}}=\frac{e^{ -\beta h\left ( 2\sum_{i=1}^{N}\sigma_i - N  \right )-\beta h}}{(d_N^{(n)}+d_N^{(n+1)}+n_N^{(n+1)})
^{\beta}}\le\frac{e^{ -\beta h\left ( 2\sum_{i=1}^{N}\sigma_i - N  \right )}}{(d_N^{(n)}+n_N^{(n+1)})
^{\beta}}e^{\beta |h|}.$$
For the lower bound we just need the $AM_{N-1}A$ terms
$$\frac{e^{ -\beta h\left ( 2\sum_{i=1}^{N}\sigma_i - N  \right )+\beta h}}{(d_N^{(n)}+n_N^{(n)}+n_N^{(n+1)})
^{\beta}}\ge\frac{e^{ -\beta h\left ( 2\sum_{i=1}^{N}\sigma_i - N  \right )}}{(d_N^{(n)}+n_N^{(n+1)})
^{\beta}}\frac{e^{-\beta |h|}}{2^{\beta}},$$
where we used the fact that $n_N^{(n)}\le d_N^{(n)}$.
Thus we get for $Z^A_{N+1}(\beta,h)=Z^{AM_{N-1}A}_{N+1}(\beta,h)+Z^{AM_{N-1}B}_{N+1}(\beta,h)$
 $$2^{-\beta}e^{-\beta |h|}Z^A_N(\beta,h)\le Z^A_{N+1}(\beta,h)\le 2e^{\beta |h|}Z^A_N(\beta,h)$$
for any $\beta \ge 0$ and $h\in {\mathbb R}.$

Finally, we prove a lemma which allows us to bound $Z_N^{B}(\beta,h)$. 
Consider a $(2\times 2)$ matrix
$M=\left({m_1\atop m_3}{m_2\atop m_4}\right)$
and define the operator $\sim$ via
$\tilde{M}:=\left({m_4\atop m_2}{m_3\atop m_1}\right)$.
Then we have the following result.
\begin{lem}\label{l1}
Let $M= AZ_1Z_2\ldots Z_N$, where $ Z_i \in \{A,B\}$,
with $A=\left({1\atop 1}{0\atop 1}\right)$ and $ B=\left({1\atop 0}{1\atop 1}\right)$.
 Then
$\tilde{M}=B\tilde Z_1\tilde Z_2\ldots 
\tilde Z_N$, i.e. the $\sim$ operator exchanges $A$ and $B$.
\end{lem}
{\it Proof.} We will use mathematical induction. It is easy to see that
${A}=\tilde B$ and $B=\tilde A$.
From matrix multiplication follows
$B\tilde{M}=\left({m_2+m_4\atop m_2}{m_1+m_3\atop m_1}\right)$ and
$AM=\left({m_1\atop m_1+m_3}{m_2\atop m_2+m_4}\right)$.

Clearly the $\sim$ operation is a 1-to-1 map of the set of all chains $AM_N$ onto $BM_N$. Furthermore, the magnetic field term in the energy of each
chain changes sign under this operation, so that the bounds just obtained for $Z_N^{A}(\beta,h)$ may be applied to $Z_N^{B}(\beta,h)$. Therefore
$$2^{-\beta}e^{-\beta |h|}\le\frac{Z_{N+1}}{Z_N}\le 2 e^{\beta |h|}.$$
Note that the proof is easily adapted to the KSC model.


\begin{thebibliography}{15}

\bibitem {K-O}
P. Kleban, and \"{O}zl\"{u}k, {\it A Farey fraction spin chain}, Commun. Math. Phys. {\bf 203}, 635-647 (1999).

\bibitem{FK}
J. Fiala, P. Kleban and A. \"{O}zl\"{u}k {\it The phase transition in statistical models defined on Farey fractions},
J. Stat. Phys. {\bf 110}, 73-86 (2003).

\bibitem{Pe}  M. Peter, {\it The limit distribution of a number-theoretic function arising from
 a problem in statistical mechanics}, J. Number Theory {\bf 90}, 265-280 (2001).

\bibitem{Ka-O} J. Kallies, A. \"Ozl\"uk, M. Peter and C. Snyder,{\it On asymptotic properties of a number
 theoretic function arising from a problem in statistical mechanics} Commun. Math. Phys. {\bf 222}, 9-43 (2001). 

\bibitem{C-Kn} P. Contucci, P. Kleban, and A. Knauf,{\it A fully magnetizing
 phase transition}, J. Stat. Phys. {\bf 97} 523-539 (1999).

\bibitem{K} A. Knauf, {\it On a ferromagnetic spin chain}, Commun. Math. Phys. {\bf 153}, 77-115 (1993).

\bibitem{C-K} P. Contucci, and A. Knauf, {\it The phase transition of the number-theoretic
 spin chain}, Forum Mathematicum {\bf 9}, 547-567 (1997).

\bibitem{Aiz1} M. Aizenman, J. T. Chayes, L. Chayes, C. M. Newman, {\it Discontinuity of the Magnetization in One-Dimensional $1/(x-y)^2$ Ising and Potts Models}, J. Stat. Phys. {\bf 50}, 1-40 (1988).

\bibitem{Aiz2} M. Aizenman, C. M. Newman, {\it Discontinuity of the Percolation Density in One-Dimensional $1/(x-y)^2$ Percolation Models}, Commun. Math. Phys. {\bf 107}, 611-647 (1986). 

\bibitem{K-o}A. Knauf, {\it The number-theoretical spin chain and the Riemann zeros},
 Commun. Math. Phys. {\bf 196}, 703-731  (1998).

\bibitem{G-K} F. Guerra and A. Knauf, {\it Free energy and correlations of the number theoretical spin chain}, J. Math. Phys. {\bf 39}, 3188-3202 (1998).

\bibitem{F} Feigenbaum, M. J., Procaccia, and T. Tel, 
{\it Scaling properties of multifractals as an eigenvalue problem}, Phys. Rev. A {\bf 39}, 5359-5372  (1989).

\bibitem{Ca}
J. Cardy {\it Scaling and Renormalization in Statistical Physics}, Cambridge University Press 1996.

\bibitem{We}
F. J. Wegner, E. K. Riedel, {\it Logarithmic Corrections to the Molecular-Field Behaviour of Critical and Tricritical Systems},
 Phys. Rev. B {\bf 7}, 248-256 (1973).

\bibitem{Diss} T. Prellberg, {\it Maps of intervals with indifferent fixed points: 
thermodynamic formalism and phase transition}, Ph.D. thesis, Virginia Tech (1991). 

\bibitem{P-S} T. Prellberg, and J. Slawny, 
{\it Maps of intervals with indifferent fixed points: 
thermodynamic formalism and phase transition}, J. Stat. Phys. {\bf 66}, 503-514 (1992).

\bibitem{Kan}
S. Kanemitsu, {\it Some sums involving Farey fractions},
Analytic number theory (Japanese) (Kyoto, 1994).
Surikaisekikenkyusho Kokyuroku {\bf No. 958}, 14-22 (1996).

\bibitem{Jap}
K. Shigeru, K. Takako and Y. Masami, {\it Some sums involving Farey fractions II.}, J. Math. Soc. Japan, {\bf 52}, 915-947 (2000). 

\bibitem{N}
J. F. Nagle {\it The one-dimensional KDP model in statistical mechanics}, Am. J. Phys. {\bf 36} (12), 1114-1117 (1968). 

\bibitem{P-s} T. Prellberg, 
{\it Complete determination of the spectrum of a transfer operator associated with intermittency}, 
preprint [arXiv: nlin.CD/0108044], (2001).

\bibitem{R} D. Ruelle, {\it Thermodynamic Formalism}, Addison-Wesley, (1978).

\bibitem{L-Z} J. Lewis and D. Zagier, {\it Period functions for Maass wave forms}, 
Annals of Mathematics {\bf 153}, 191-258 (2001).

\bibitem{M} D. Mayer, {\it The thermodynamic formalism approach to Selberg's zeta 
function for PSL(2,$\mathbb Z$)}, Bull. AMS {\bf 25}, 55-60 (1991).



\end{thebibliography}
\end{document}